# Electrically-Driven Two-Dimensional Semiconductor Microcavity Laser

Zheng-Zhe Chen[1,2,4]†, Hsiang-Ting Lin[1]†, Chiao-Yun Chang[1,3], Adil Muhammad[1,5], Po-Cheng Tsai [1,6], Tsung Sheng Kao[5], Chi Chen[1], Shu-Wei Chang[1,5], Shih-Yen Lin[1,6], and Min-Hsiung Shih[1,5,7]∗

[1]Research Center for Applied Sciences (RCAS), Academia Sinica, Taipei 11529, Taiwan.

[2]Department of Physics, National Taiwan University, Taipei 10617, Taiwan.

[3]Department of Electrical Engineering, National Taiwan Ocean University, Keelung 20224, Taiwan.

[4]Nano Science and Technology, Taiwan International Graduate Program, Academia Sinica and National Taiwan University, Taipei 11529, Taiwan.

[5]Department of Photonics and Institute of Electro-Optical Engineering, National Yang Ming Chiao Tung University, Hsinchu 30010, Taiwan.

[6]Graduate Institute of Electronics Engineering, National Taiwan University, Taipei 10617, Taiwan.

[7]Department of Photonics, National Sun Yat-sen University, Kaohsiung 80424, Taiwan.

†These authors contributed equally to this work.

*Corresponding author. Email: mhshih@gate.sinica.edu.tw

**Abstract**

Two-dimensional (2-D) monolayer transition-metal dichalcogenides (TMDCs) are promising materials for realizing ultracompact, low-threshold semiconductor lasers. And the development of the electrical-driven TMDC devices is crucial for enhancing the integration potential of practical optoelectronic systems. However, at current stage, the electrically-driven 2-D TMDC laser has never been realized. Herein, we have developed the first electrically-driven 2-D TMDC microcavity laser. In this device, an alternating current (AC) generates electroluminescence lasing in suspended monolayer $WSe_2$ integrated on a microdisk cavity. We also analyzed the input–output curve, bandwidth narrowing, and second-order coherence to confirm the lasing characteristics in room temperature. Our realization of the room temperature AC-driven 2-D TMDC laser establishes a new area of research on electrically pumped compact lasers and is likely to assist with the implementation of diverse TMDC-based practical photonic devices in the future.

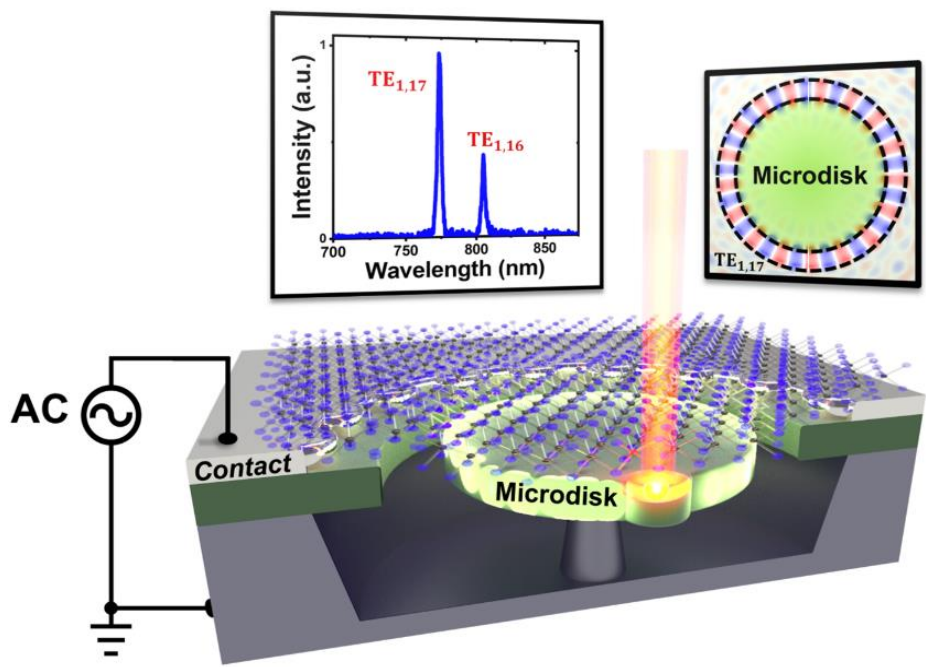

Monolayer transition-metal dichalcogenides (TMDCs) hold considerable potential in nanophotonics field thank to their exceptional properties, including ultra-thin dimensions, direct bandgap structures, and high exciton-binding energy[1-6]. Recently, researches based on TMDC-based electroluminescent (EL) has been widely studied and led to diverse achievements[7-17]. One of strategies for constructing TMDC EL devices involves integrating TMDCs with a dielectric layer to create capacitor-like structures[18-21]. In this configuration, the device is operated by alternating current (AC), and transient carrier tunneling from the capacitor component to the TMDC is used to generate pulsed EL emission. The architecture of the AC-driven TMDC EL device has a simplified structure that can be designed over large areas by using chemical vapor deposition (CVD)-grown TMDC material[18-20]. The innovative development of these devices has established a new direction in the exploration for creating practical and scalable TMDC-based EL devices[18].

The unique properties of TMDCs indicate that they can be used as the gain medium in lasing devices[22-31], which has led to TMDCs attracting considerable attention in research. They are particularly suitable for use in ultra-compact and low-threshold lasers[22,30], the application of which have led to breakthroughs in various fields. For most of the TMDC laser, the gain material TMDC was directly attached to the cavity substrate to strengthen the coupling between cavity mode and gain[23-26,29]. However, when TMDCs are placed on substrates, several problems emerge, including the shielding effect, the presence of surface states, and defects at the substrate interface[32-34]. And one of the simple solutions to address these challenges is by suspending materials. Without direct attach to substrate, suspended TMDCs are not subject to substrate-induced strain, defects, or impurities[32,35-37]. This, in turn, substantially enhances their light emission capabilities and the overall emission quantum yield in light emission applications[32,37,38]. These advantages collectively add to the potential of suspended TMDC materials as a promising lasing gain medium for more efficient and reliable laser devices.

Although the use of TMDCs in conjunction with various micro-nano-cavities for optical-pumped lasing devices has been extensively studied[22-31], explorations of electrically driven TMDC-based lasers remain limited. Herein, we present a novel approach that involves integrating suspended tungsten diselenide ($WSe_2$) with gap microdisk cavity and use AC injection to achieve electrically driven TMDC lasers in room temperature. We have further explored the electrically driven lasing characteristics and corresponding coherent properties to confirm our laser performance.

**Electrically pumping monolayered WSe₂ luminescence and laser device**

Fig. 1a presents a schematic of the developed electrically pumped laser that comprises a suspended monolayer WSe$_2$ flake, a silicon-nitride (Si$_3$N$_4$) microdisk cavity, a heavily doped silicon (p$^{++}$-Si) bottom electrode, and a surrounding silver (Ag) gear top electrode. The monolayer WSe$_2$ flake is suspended between the microdisk cavity and top electrode, and an AC square wave signal is applied as the electrical injection[18-21]. The microdisk structure plays multiple roles in this electrically driven lasing device. The Si$_3$N$_4$ microdisk cavity provides high-quality whispering gallery modes (WGMs) with a high Q factor for lasing resonance[25,27,28.30,31]. In addition, the microdisk structure acts as a dielectric capacitor for AC electrical injection. The center Si pillar supporting the microdisk functions as the bottom gate electrode. Furthermore, the inclusion of a nanometer-scale scatterer (nanoscatterer) positioned near the edge of the microdisk helps facilitate the vertical free-space EL signal collection[28]. The surrounding top electrode was designed in a gear shape to further intensify the EL from the suspended WSe$_2$. Demonstrated in Supplementary Fig. S1, by comparing the intensity of monolayered WSe$_2$ EL spectra, the intensity pumped by gear electrode exhibited an-order stronger than the one in normal electrode. There are two possible explanations: one is the increasement of total length of contact-material interface, and the other is that the existence of gear may help extract material's luminescence from lateral to vertical and further collected. The detailed fabrication process is described in the Methods section.

Fig. 1b displays a scanning electron microscopy (SEM) image of the realized laser device. The suspended monolayer WSe$_2$ flake, which served as a two-dimensional gain medium, was suspended between the microdisk (with a diameter around 3.7 μm) and surrounding top electrode, which were separated by a narrow, 400-nm air gap. We confirmed the monolayer structure of WSe$_2$ through Raman analysis (Supplementary Fig. S2). The Raman spectrum with characteristic E$^1_{2g}$, A$_{1g}$ peaks but absence of interlayer B$^1_{2g}$ peak at near 310 cm$^{-1}$ wavenumber indicates the monolayer property of WSe$_2$ material[39-41]. The atomically thin WSe$_2$ monolayer was securely suspended on the narrow gap without breakage occurring (yellow region in Fig. 1b inset). In addition, the WGMs of the microdisk cavity were surrounded by the narrow air gap, which led to a higher proportion of cavity mode being distributed in the gap site (Fig. 4c), facilitating the coupling strength between the cavity resonances and suspended WSe$_2$ EL. For suspended WSe$_2$ monolayers, environmental effects, such as those from the bottom substrate, can be effectively isolated[32-34]. To confirm the effect of suspending the material, we compared the EL and photoluminescence (PL) between suspended and non-suspended (i.e., in contact with Si$_3$N$_4$ substrate) WSe$_2$ in our devices. According to the EL (Fig. 1c) and PL (Supplementary Fig. S3) spectra, the exciton-dominated emission intensity of the suspended material was approximately five times stronger than that of the contacted material under the same pumping conditions[32,37,38]. This finding supported the potential of the suspended monolayer WSe$_2$ as a lasing gain medium for our electrically driven laser.

Our electrically driven laser operates by injecting an AC square wave voltage signal[18-21]. Lasing emission is achieved through the induced transient EL in WSe$_2$ coupled to cavity WGM resonances. Fig. 1d presents a schematic illustrating the process of electrically driven transient EL emission from the monolayer WSe$_2$[18,19]. In our device, the Si$_3$N$_4$ microdisk serves as the capacitive component between the gate electrode and gain medium WSe$_2$. When a bias voltage (V) is applied, it generates a strong electric field across the Si$_3$N$_4$ capacitor layer and WSe$_2$ as the square wave V signal switches its polarity. In this process, marked as step 1, when a steady positive bias voltage is applied to the back gate, it blocks holes in the WSe$_2$ due to the Schottky barrier. The polarity of V instantaneously changes from positive to negative (marked as step 2), and the capacitive layer, which is $Si_3N_4$ in this case, cannot respond to the fast field change as it occurs. This leads to transient band bending near the Schottky contact resulting in substantial transient carrier tunneling between the $WSe_2$ and contact. The tunneling electrons from the contact then recombine with holes accumulated in $WSe_2$ during step 1, generating radiation dominated by excitons at the bandgap. Subsequently, similar processes occur during step 3 and 4. Therefore, pulsed EL can be efficiently generated at steps 2 and 4, providing an opportunity for further coupling to the microdisk cavity and enabling realization of AC-driven $WSe_2$ laser.

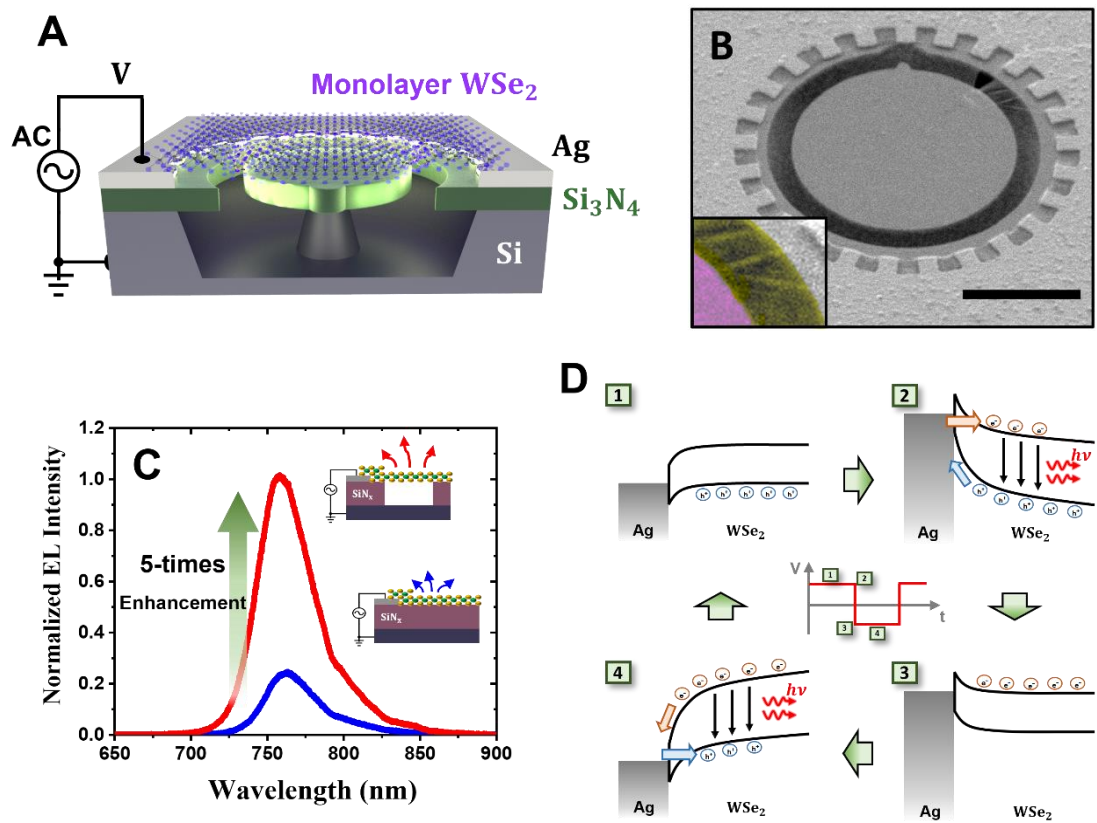

**Fig.1|Schematic of laser device and suspended WSe₂ EL.**
**a** Schematic of the electrically AC-driven WSe$_2$ laser. Monolayer WSe$_2$ is positioned on the top of the device, suspended between a silicon-nitride microdisk cavity and silver electrode. **b** Scanning electron microscopy image of the fabricated laser device. The microdisk diameter and air gap width are 3.7 μm and 400 nm, respectively. Scale bar is 2 μm. (Inset: yellow region represents monolayer WSe$_2$ suspended on the air gap; magenta region indicates the microdisk) **c** EL spectra of suspended and contacted monolayer WSe$_2$ under the same AC pumping condition (16 Vpp and 14 MHz). Suspended WSe$_2$ EL signals are collected at the center of the air gap, which is at a distance of approximately 400 nm from the electrode edge. Contacted WSe$_2$ EL signals are collected at the same distance from the electrode edge to minimize differences in carrier density distribution. **d** Schematic of band diagram at the Ag–WSe$_2$ interface illustrating four steps of applying different voltage polarities.

**Electrically pumping WSe$_2$ lasing characteristics**

All the electrically driven lasing emissions were characterized at room temperature. An AC square wave voltage signal was applied between the top and back electrode for excitation. EL lasing signals were collected using a 100x objective lens and analyzed in free space by using a spectrograph and charge-coupled device detector. To collect data on the lasing emission resulting from the transient EL in the WSe$_2$ coupled with WGMs, the objective lens was focused on the nanoscatterer region of the microdisk. Details regarding the optical characterization methods are provided in the Materials and Methods section. When the AC voltage signal injection was sufficient, several sharp lasing peaks and spontaneous emissions occurred (Fig. 2a). Two strong lasing peaks appeared at approximately 773 and 805 nm and were associated with the WGMs of TE$_{1,17}$ and TE$_{1,16}$ (with radial mode number n = 1 and azimuthal mode numbers m = 17 and 16), respectively. Under the same pumping density, the lasing signal exhibited a much higher peak intensity than spontaneous EL did. In this study, we mainly focused on the most notable, 773-nm lasing peak in our investigation of lasing properties such as intensity, linewidth, and coherence. EL lasing spectra with different pumping voltages, ranging from 12 to 19 V, are presented in Fig. 2b. At a low excitation level, only weak EL was detected from the WSe$_2$. As the injection voltage increased, the lasing resonances resulting from the coupling between the WSe$_2$ EL and microdisk WGMs became more notable, and the lasing peak signal became more evident when a higher voltage was injected.

Fig. 2c-d display the input–output curves of the 773-nm lasing peak and spontaneous EL intensity, respectively, as the function of the applied bias voltage (V) or the corresponding injected carrier density (cm$^{-2}$). Only one turning point was observed in the spontaneous EL turn-on curve (Fig. 2d), whereas two significant

thresholds were noted in the lasing turn-on curves (Fig. 2c). The first threshold, occurring at approximately 12 V, represents the point at which the carriers began to recombine and emit spontaneous EL through efficient Schottky barrier tunneling, as indicated in the lasing and spontaneous EL curves. The voltage ($V_{thr-SB}$) of this threshold was mainly determined by the material bandgap and device resistance. The other significant threshold ($V_{thr-ST}$), which was observed in only the lasing curve at approximately 14 V, resulted from the emergence of stimulated emission. These results from the coupling between the WSe$_2$ EL and microdisk cavity provide evidence of lasing turn-on. Furthermore, a typical lasing linewidth narrowing when a higher voltage was injected was observed (Fig. 2e). The full width at half maximum (FWHM) of the lasing peak decreased from 4.5 to 2.8 nm, which is consistent with the threshold findings presented in Fig. 2c.

To gain a deeper understand of the lasing performance of our electrically pumped WSe$_2$ laser, we estimated the injected carrier density ($N_{in}$) from the applied voltage by using the following formula[18,20]:

$$Injected\ carrier\ density\ (N_{in}) = \frac{C_g(V_{pp}-\frac{\varphi_{Bn}+\varphi_{Bp}}{q})}{q},$$

where $C_g$ is the gate capacitance ($39.4\ nF\ cm^{-2}$); $V_{pp}$ is the applied voltage (V); q is the unit charge; and $\varphi_{Bn} + \varphi_{Bp}$ is the sum of the barrier heights for electrons and holes, which is equivalent to the bandgap energy $E_g$ of the material ($\varphi_{Bn} + \varphi_{Bp} = E_g$). As illustrated in Fig. 2f, the EL lasing intensity was replotted as a function of the injected carrier density, N, by using a double logarithmic scale. Intensity thresholds occurred at approximately $N_{in}$ = 2.5E+12 (cm$^{-2}$) and $N_{in}$ = 3.4E+12 (cm$^{-2}$), respectively representing the points at which the voltage/carrier density was sufficiently high to achieve significant tunneling through the Schottky barrier and at which the generated photon number became sufficiently large for population inversion and stimulated emission. Within the green region of Fig. 2f (2.5E+12 cm$^{-2}$ < $N_{in}$ < 3.4E+12 cm$^{-2}$), spontaneous emission (SPE) predominantly governed the EL radiation process. As the applied voltage increased, the carrier density surpassed 3.4E+12 cm$^{-2}$ (blue region), and the slope of EL intensity as a function of $N_{in}$ changed, indicating the onset of stimulated emission (STE). Collecting a pure STE signal from CVD-grown TMDC can be challenging because of the presence of intrinsic material defects and low quantum efficiency. However, the ratio of STE is believed to continue increasing as N increases until saturation is reached, which is typical for semiconductor lasers. In the present study, to evaluate the spontaneous emission coupling factor (*β*) of our electrically pumped WSe$_2$ microdisk laser, we applied a conventional rate equation model to achieve reasonable fitting. We observed that *β* = 0.18 exhibited the best fit to the data, indicating that approximately 18% of the spontaneous emission was coupled into the cavity mode for STE generation. According to these fitting results, other significant parameters, including material's gain (g), cavity confinement factor (Γ), effective mode volume ($V_m$), and photon lifetime ($\tau_p$), were able to be estimated on the basis of our experimental or simulation results (Supplementary S4, S5). By implying further PL measurements, we estimated our material threshold gain is around 7455 cm$^{-1}$, which is quite compatible to previous WGM-based TMDC laser literatures[27,30]. And also, based on the PL results, the pumping threshold is at 77 μW, which has met the quantum threshold condition we estimated, showing that our lasing device can satisfy the real lasing phenomenon experimentally and theoretically (Supplementary Fig. S5).

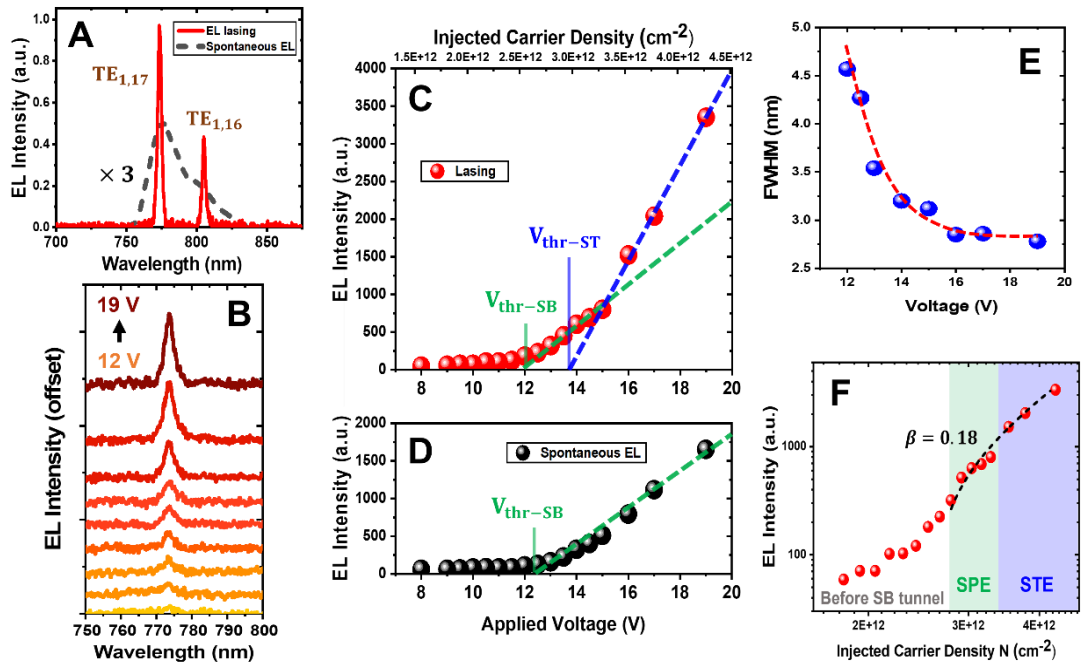

**Fig.2|Lasing characteristics.**
**a** EL lasing (red) and spontaneous EL (gray) spectra of monolayered $WSe_2$ on a microdisk with AC square wave operating at 19 V and 16 MHz at room temperature. **b** EL lasing spectra under pumping voltages ranging from 12 to 19 V. **c, d** Input–output curves of intensity extracted from lasing and spontaneous EL spectra as a function of applied voltage (or carrier density). Green and blue dashed lines represent the intensity of spontaneous and STEs, respectively. Their turn-on voltage thresholds are marked as $V_{thr-SB}$ and $V_{thr-ST}$. **e** FWHM of the 773-nm lasing peak as a function of applied voltage. **f** Plot of the lasing intensity as a function of injected carrier density $N_{in}$ in the double logarithmic scale. White, green, and blue regions respectively represent the points at which carriers are insufficient for efficient recombination and at which spontaneous and stimulated EL emission dominate the radiative process. Black dashed line represents the fitted rate equation for spontaneous emission coupling factor $\beta = 0.18$.

**Second-order coherence characteristics**

To further characterize the coherent behavior of our WSe$_2$ on the microdisk cavity, we used a Hanbury Brown-Twiss (HB-T) interferometer setup for measuring the corresponding second-order photon correlation function $g^{(2)}(\tau)$[28,42-45]. This function is expressed as follows:

$$g^{(2)}(\tau) = \frac{\langle n(t)n(t+\tau)\rangle}{\langle n(t)\rangle\langle n(t+\tau)\rangle}$$

where $n(t)$ is the time-dependent photon number. The schematic of the measuring setup is illustrated in Supplementary Fig. S6. Fig. 3 display the measured correlation function $g^{(2)}(\tau)$ as a function of delay time for the EL signal both below (13 and 14 V) and above (16 and 18 V) the lasing threshold condition. As indicated in Fig. 3a and 3B, below the threshold, the EL signal exhibits Plank's thermal (or super-Poisson) distribution, with $g^{(2)}(0)$ = 1.487 and 1.392, respectively. This finding indicates the occurrence of the photon bunching behavior of spontaneous emission. The STE exhibited a Poisson statistic distribution, where $g^{(2)}(\tau)$ remained close to 1 across all delay times $\tau$, confirming typical coherent laser performance. As presented in Fig. 3c and 3D, as the applied voltage approached and exceeded the threshold voltage, the value of $g^{(2)}(0)$ decreased to 1.032 and 1.004, respectively, demonstrating typical coherent lasing performance. A similar phenomenon was observed when the device was pumped with a 532-nm laser (Supplementary Fig. S6). To enable comparison and illustrate the observed trend, $g^{(2)}(0)$ as a function of the applied voltage is plotted in Fig. 3e. These results obtained from the second-order photon correlation function measurements provide further evidence of the successful realization of an electrically pumped TMDC laser by using our WSe$_2$ microdisk cavity.

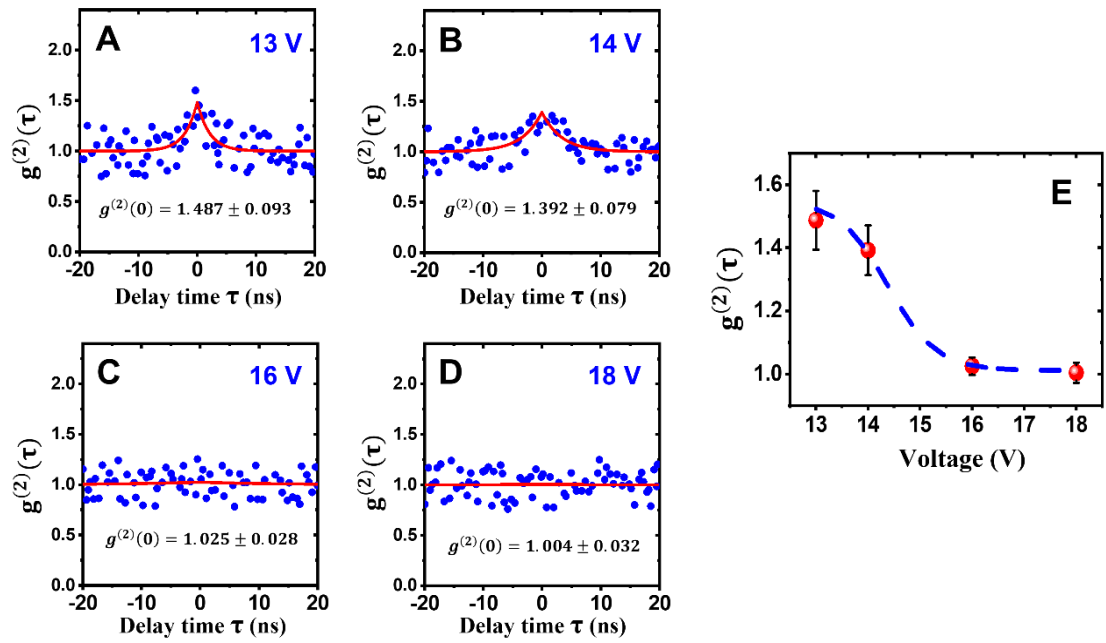

**Fig.3|Second-order coherence characteristics.**

**a–d** Second-order coherence function measurements of WSe$_2$ EL on a microdisk device with 13, 14, 16, and 18 V of pumping voltage. Blue dots represent raw data, and red solid lines represent corresponding fitted results from the Siegert relation. **e** Value of the second-order coherence function at zero delay time $g^{(2)}(0)$ in the function of applied voltage. Blue dashed curve is fitted to indicate the tendency.

**Resonant mode and intensity distribution engineering**

Fig. 4b presents the experimental EL intensity top-view profile. The green and blue dashed circles represent the edge of the microdisk cavity and top contact, respectively. The two-dimensional profile revealed that the intensity was mainly distributed over the suspended $WSe_2$ region, which exhibited considerable spatial overlap with the gap microdisk's WGM (Fig. 4c). In addition, a stronger EL intensity was observed at the nanoscatterer location (area within the white square in Fig. 4b), confirming the importance of the scatterer for efficiently extracting lateral radiation for vertical collection through the lens[28]. Fig. 4c depicts the top-view mode profile from the WGM of the magnetic field ($H_z$) of our monolayered $WSe_2$ on the gap microdisk, which was obtained using the finite element method (FEM). The black circles represent the inner and outer edges of the air gap, where the cavity WGM was mainly distributed. This distribution spatially corresponded to our experimental results (Fig. 4b). The cross-section mode profile near the air gap is depicted in Fig. 4d, illustrating the strong overlap between the WGM and the gain material $WSe_2$, which provides efficient coupling and further enhances lasing generation.

To spatially optimize the EL intensity of the $WSe_2$ for coupling with the microdisk cavity, we investigated two crucial parameters: electrode–gap distance (D) and gap width (G) (Fig. 4a). Our previous study revealed the importance of suspension for enhancing the carrier recombination rate. This finding indicates that controlling the precise location where carriers meet, recombine, and emit within the gap is essential.

Determining the electrode–gap distance, D, is critical to precisely control the carriers' recombination at the air gap site. We fixed the gap width at 400 nm and calculated the optimal D to be approximately 383 nm by using impedance calculations between our device's effective RC circuit and the AC pumping conditions (Supplementary Fig. S7). This result was experimentally verified (Fig. 4e). The measured EL intensity distribution profile along the X direction for D when its value ranged from 100 to 1200 nm is plotted in Fig. 4e. In each figure, the gray area marks the position of the air gap. The profiles were compared, and the group with D = 400 nm (blue solid line) exhibited the most favorable gap-intensity spatial overlap and strongest EL. In other words, the carriers most efficiently met and recombined at the gap site when D was approximately 400 nm. The right part of Fig. 4e depicted the schematic diagram illustrating relative position of air gap and where carriers predicted to meet along X direction with different D.

Due to the affections of carrier diffusion and material strain, it is also critical to optimize the effective suspending area. In the current study, we optimized the EL intensity by finely tuning the gap width (G). As illustrated in Fig. 4f, with D fixed at 400 nm, we varied gap width from 0 to 1200 nm. The schematic diagrams of relative air gap size and where carriers meet along X direction with different G are also depicted in the right part. On the basis of our measurement results, we plotted the normalized EL intensity as a function of G in Fig. 4g. When G was smaller than 300 nm, the effect of suspension was nonsignificant. The EL intensity reached its maximum at G ranged from 300 to 400 nm and started to decrease markedly with larger gap width, possibly due to tensile strain on materials.

These findings demonstrate not only where carriers may recombine but also the optimal gap size for generating strongest EL to couple with the cavity mode, indicating that these D and G settings achieved the optimized AC-driven lasing behavior at room temperature.

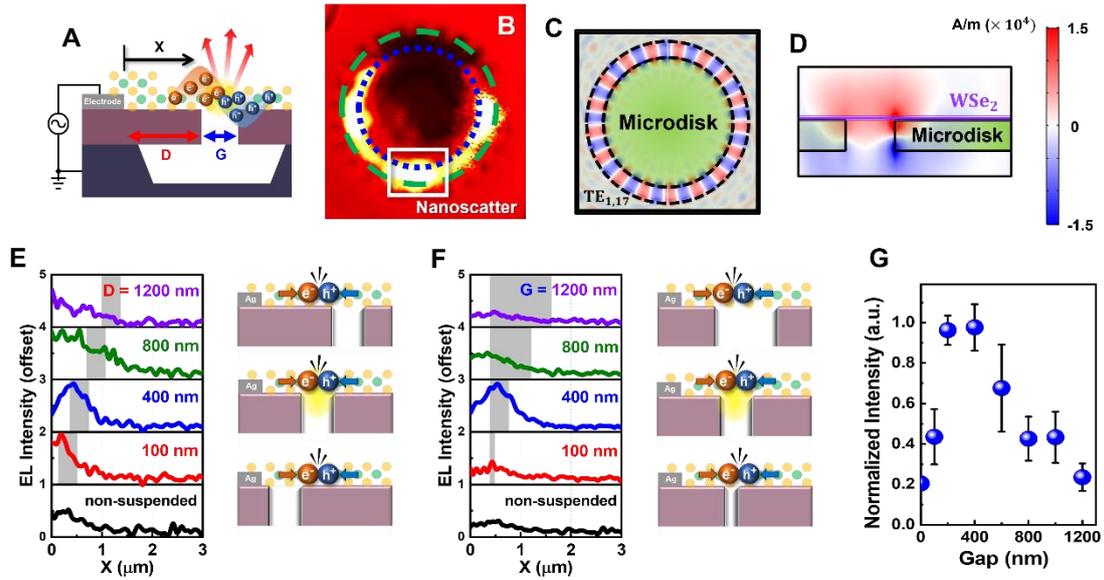

**Fig.4|Resonant mode and intensity distribution profile.**

**a** Schematic of the cross-section of our device, with D (contact-gap distance) and G (contact-gap gap width) marked. **b** Two-dimensional EL intensity profile of our WSe$_2$ on a microdisk laser under 18 V$_{pp}$ and 16-MHz AC pumping. Green and blue dashed circles respectively indicate the outer and inner edges of the microdisk and top electrode. White square indicates the location of the nanoscatterer. **c** and **d** Top-view and side-view mode profiles of horizontal magnetic field (H$_z$) distribution from monolayer WSe$_2$ on a gap microdisk, respectively. Black solid lines represent SiN$_x$ edge. **e** and **f** EL intensity distribution profiles along the X direction away from the electrode edge with different D and G tunings, respectively. Gray areas represent where the gap spatially overlap. Right schematics depict the relative position of the air gap and where the carriers meet along the X position with different D and G. **g** Plot of the normalized EL intensity as a function of the air gap width.

**Conclusion**

By integrating suspended monolayered WSe$_2$ with a Si$_3$N$_4$ microdisk, we successfully fabricated the first electrically AC-driven monolayered TMDC laser operable at room temperature. Our realizations of electrically driven TMDC-based lasers would not only potentially enable creation of compact and efficient optoelectronic devices but also advance the field of integrated photonics, enabling the development of practical integrated photonic systems.

## Methods

### Device Fabrication

First, a 150 nm-thick $SiN_x$ layer was deposited on a highly-doping $p^{++}$ silicon substrate through low pressure chemical vapor deposition (LPCVD). Next, to fabricate the microdisk cavity, we utilized electron beam lithography (EBL) to creat the pattern. The electron beam resist PMMA was spin-coated on $SiN_x$ for EBL development. After development, the substrate then went under an inductively coupled plasma reactive-ion etching (ICP-RIE) process. The residual resist PMMA was removed with oxygen plasma in the same ICP-RIE. Next the as-etched substrate was immersed in 40 wt% KOH solution at 60 °C to form a suspended microdisk structure through an anisotropic wet etching of the Si substrate. Hereafter, a second time aligned EBL development was applied for top contact. Then, the 50 nm silver electrode is accomplished by thermal deposition and lift-off process. Finally, CVD-grown monolayered $WSe_2$ was further wet transferred on the substrate[46], suspended between microdisk cavity and top metal contact.

### EL and PL Measurement

All the EL and PL spectrum were measured by the home-build confocal-microscope μ-PL system. Only spectrum Fig. 2a was collected through multi-track mode, which is filtered by grating in our charge-coupled device (CCD) to extract the optical signal for longer wavelength part, helped emphasize the $WSe_2$ EL lasing peak's signal-noise ratio. Otherwise, other spectra were collected full wavelength range, without any external filter. For lasing spectra in Fig. 2a-b, they were extracted from the raw spectra data, where the spontaneous emission part can be eliminated with reasonable exciton, trion and defect mode fittings.

For EL measurements, an AC function generator is applied to create square wave at MHz-scale frequency, then the electrical signal and ground is contacted to the top electrode and back gate, respectively. And for PL measurements, a 532 nm wavelength CW diode laser is utilized as source (except the time-resolved PL measurement) and focused by a 100×, N.A.= 0.7 objective lens with spot size around 1 μm. And the PL radiation is collected by the same objective lens and further recorded/analyzed by a spectrograph with thermoelectrically cooled CCD. The resolution of spectrum was approximately 0.06 nm. For long-time measurements, like time-resolved PL (TRPL) or second-order coherence, the signal was collected using a single photon avalanche diode connected with a time-corrected single photon counting (TCSPC) module.

**Acknowledgments:**

This work was supported by the Innovative Materials and Analytical Technology Exploration (i-MATE) program of Academia Sinica in Taiwan, the National Science and Technology Council (NSTC) and the Ministry of Science and Technology (MOST) in Taiwan under Contract No. NSTC 112-2112-M-001- 074, MOST 111-2112-M-001-078, and MOST 110-2112-M-001-053.

# Supplementary Information

**S1. WSe$_2$ EL pumped with normal and gear electrode**

By comparing the intensity of monolayered WSe$_2$ EL spectra, the intensity pumped by gear electrode exhibited an-order stronger than the one in normal electrode. There are two possible explanations: one is the increasement of total length of contact-material interface, and the other is that the existence of gear may help extract material's luminescence from lateral to vertical and further collected.

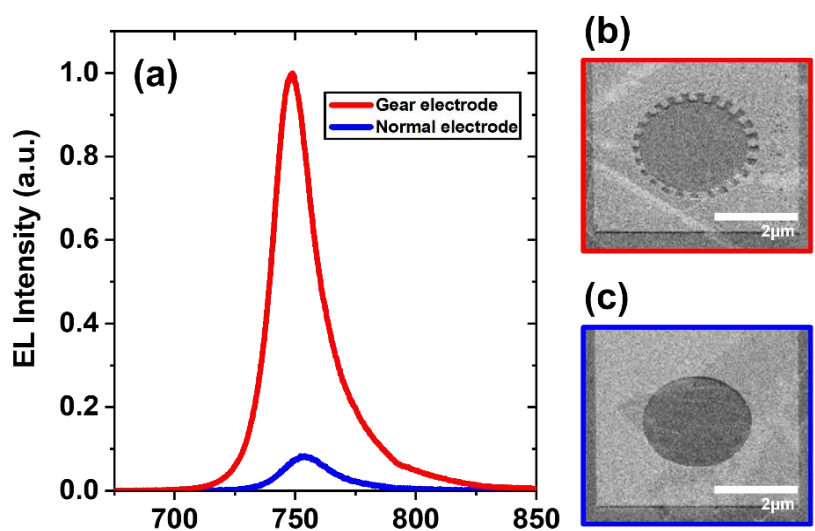

**Fig. S1 a**. Normalized EL spectra from WSe$_2$ pumped by gear (red) and normal (blue) top contact under 16 Vpp, 16 MHz AC condition. **b, c.** SEM pictures of monolayered WSe$_2$ contacted with gear (red) and normal (blue) electrodes. Scale bar is 2 μm.

## S2. Raman spectrum of the monolayered WSe$_2$

The Raman spectrum with characteristic E$^1_{2g}$, A$_{1g}$ peaks but absence of interlayer B$^1_{2g}$ peak at near 310 cm$^{-1}$ wavenumber indicates the monolayer property of WSe$_2$ material. A 532 nm laser was applied as the pumping source.

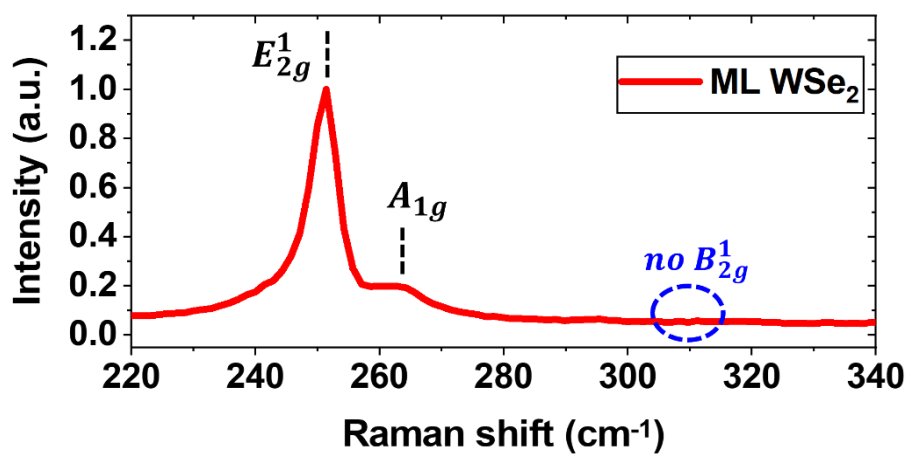

**Fig. S2** Raman spectrum of monolayer WSe$_2$ on the device.

## S3. PL spectra of contacted and suspended monolayered WSe$_2$

Being similar to the difference of EL spectra intensity, the PL spectra also exhibited about 5 times intensity enhancement after suspending the material.

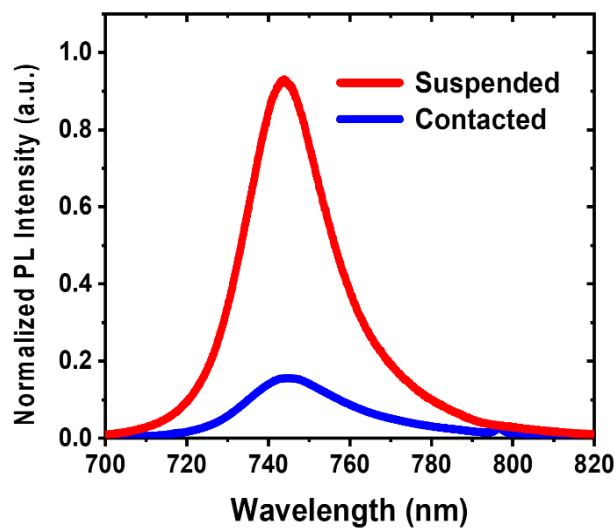

**Fig. S3** Normalized PL spectra of contacted (blue) and suspended (red) WSe$_2$. Pump source: 50 μW, 532nm CW laser.

## S4. The lasing characteristics of the WSe₂ on microdisk cavity

The spontaneous-emission coupling factor $\beta$ is the fraction of spontaneous emissions coupled into a cavity mode and determines the turn-on behavior of the laser around threshold. The steady-state rate equations were utilized to fit the experimental input-output curve (Fig. 2f) to extract $\beta$. The relation between carrier density ($N$) and photon density ($P$) is described using the following equation (Eq. S1, S2), where $N_{in}$, $\hbar\omega$, and $V$ represent the injected carrier density, photon energy of pumping source and volume of the active region, respectively; $\tau$ is the radiative recombination time; $\tau_p$ is the photon lifetime; $v_g$ is the group velocity; $g(N)= a(N\text{-}N_{th})$ is the material gain; and $N_{tr}$ and $a$ represent the transparency carrier density and differential gain, respectively. Here we simply take the lifetime $\tau$ as a constant (4.77 ns) for simplicity, measured by time-resolved PL (TRPL) with a 450 nm-pulse laser excitation in room temperature (Fig. S4). At steady the state, we set $dN/dt = dP/dt = 0$, eliminated the variable $N$, and derived a quadratic equation of $P$ as following equation (Eq. S3), where we have defined a new parameter $\bar{a} \equiv av_g\tau/\beta$; and the term proportion to $N_{tr}$ was dropped due to its smallness as compared with other terms in most of the cases. We then used Eq. S3 as the fitting function. By carefully adjusting $\bar{a}$ and $\beta$, we attained the spontaneous-emission coupling factor $\beta = 0.18$ as the best fit to our EL lasing performance.

$$\frac{dN}{dt} = \frac{\eta}{V}\frac{N_{in}}{\hbar\omega} - \frac{N}{\tau} - v_g g(N)P \qquad \text{(Eq. S1)}$$

$$\frac{dP}{dt} = \Gamma v_g g(N)P + \Gamma\beta\frac{N}{\tau} - \frac{P}{\tau_p} \qquad \text{(Eq. S2)}$$

$$P \approx \frac{1}{2\bar{a}\beta}\left[-\left(\bar{a}\beta\Gamma\tau_p\frac{\eta}{V}\frac{N_{in}}{\hbar\omega} - 1\right) + \sqrt{4\bar{a}\beta^2\Gamma\tau_p\frac{\eta}{V}\frac{N_{in}}{\hbar\omega} + \left(\bar{a}\beta\Gamma\tau_p\frac{\eta}{V}\frac{N_{in}}{\hbar\omega} - 1\right)^2}\right] \qquad \text{(Eq. S3)}$$

Furthermore, we used the finite element method (FEM) to calculate the eigenmodes with a frequency around the lasing peak in the microdisk cavity. We set the bulk $SiN_x$ as the cavity layer, whose refractive index n at $WSe_2$'s emission wavelength and thickness were set to be 1.8 and 150 nm, respectively. The monolayered $WSe_2$ was approximated as a thin layer with a thickness of 0.75 nm and its refractive index were determined according to a past study (complex index n+ik ≈ 4.65+0.15i around the lasing wavelength). The effective index $n_{eff}$ of our microdisk cavity was 1.59, which is lower than that of regular microdisk (1.65), indicating that the cavity mode distributes higher ratio in the free space (air) than usual.

We utilized the following equation (Eq. S4) of the effective mode volume ($V_m$) to estimate the size of cavity mode.

$$V_m = \frac{\int d\mathbf{r}\, n^2(\mathbf{r})|\mathbf{E}(\mathbf{r})|^2}{\text{Max}\,[n^2(\mathbf{r})|\mathbf{E}(\mathbf{r})|^2]} \quad \text{(Eq. S4)}$$

where n(r) is the space-dependent is refractive index and E(r) is the electric field. By calculating the ratio between the volume integral and the maximal field intensity in the microdisk, we obtained the effective mode volume $V_m$ = 0.26 µm³ = 3.33 $(\lambda/n)^3$, where λ is the lasing wavelength, and n is the refractive index of $SiN_x$.

The confinement factor (Γ) describes the overlap between the electric field and the active medium. We used the following equation (Eq. S5) to calculate it.

$$\Gamma = \frac{\int_\Omega d\mathbf{r}\, n_{WSe2}^2|\mathbf{E}(\mathbf{r})|^2}{\int d\mathbf{r}\, n^2(\mathbf{r})|\mathbf{E}(\mathbf{r})|^2} \quad \text{(Eq. S5)}$$

Herein, $n_{WSe2}$ is the refractive index of monolayered $WSe_2$ and Ω is the region of active medium. With the FEM calculation, the confinement factor Γ of the our microdisk cavity was approximately 3.32 %.

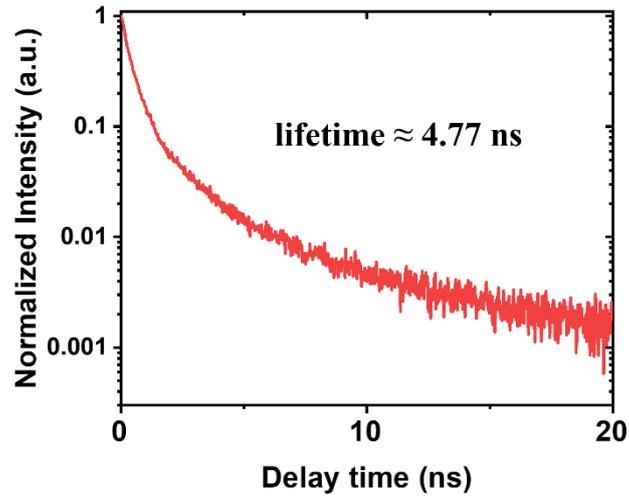

**Fig. S4** Time-resolved PL measurement to the monolayered WSe$_2$ material under 450 nm pulsed laser excitation (average power: 5 μW, repetition rate:10 MHz). The photon lifetime is estimated to be about 4.77 ns by exponential fitting.

## S5. Threshold gain and quantum threshold condition estimation by PL measurement

Based on the numerical results above, material's threshold gain is further estimated through photo-pumping experiments. Photoluminescence spectrum of our suspended WSe$_2$/microdisk laser pumped by 532 nm laser is exhibited in Fig. S5a. By gradually increasing pumping power, the $TE_{1,17}$ mode lasing peak shows up at 770 nm (Other corresponded cavity modes with different azimuthal number also reveal on spectra). By extracting integrated PL intensity from $TE_{1,17}$ and $TE_{1,16}$ mode lasing peak, the input-output curve indicated that the threshold power density $P_{th}$ is around 77 μW (Fig. S5b). A typical TRPL measurements are also taken to confirm the couple effect of microdisk cavity (Fig. S5c). Under same pulse power and repetition rate of excitation, PL emission of WSe$_2$ revealed a clear difference of intensity decay tendency between on cavity and on bare substrate. The shorter PL lifetime of WSe$_2$ on cavity than off cavity is results from the light-matter interaction between gain material and microdisk cavity.

To estimate the material's threshold gain ($g_{th}$) of our laser device, the microdisk cavity loss should be first determined. For WGM cavity, the reduction of energy can be written as below.

$$I = I_0 e^{-\frac{2\pi n_{eff}}{Q\lambda}L_r} \qquad (Eq.\ S6)$$

where n$_{eff}$ is the effective index, L$_r$ is one round-trip optical path length, Q and $\lambda$ represent the quality factor and wavelength of the cavity mode.

Thus, from Eq. S6, we can determine the total cavity loss $\delta_r$ as,

$$\delta_r = \frac{2\pi n_{eff}}{Q\lambda} \qquad (Eq.\ S7)$$

While meeting lasing threshold, the cavity loss will be compensated by the material gain, which leads to

$$\delta_r = \Gamma g_{th} \quad \text{(Eq. S8)}$$

giving the threshold gain ($g_{th}$)

$$g_{th} = \frac{2\pi n_{eff}}{\Gamma Q \lambda} = 7455 \; cm^{-1} \quad \text{(Eq. S9)}$$

where $n_{eff}$ = 1.59, $\Gamma$ = 3.32%, Q = 523.8 and $\lambda$ = 770 nm

The estimated value is also quite compatible with previous WGM-based TMDC laser reports.

Furthermore, we try to estimate if our pumping condition meet the quantum threshold condition. Experimentally, to satisfy quantum threshold condition, the pumping generation rate at threshold ($G_{th}$) should be larger than the rate of photon loss in cavity ($\kappa_{cavity}$).

$$G_{th} \geq \kappa_{cavity} \quad \text{(Eq. S10)}$$

$$\frac{P_{th}\alpha}{h\upsilon_{pump}} \geq \frac{2\pi\upsilon_{lasing}}{Q} \quad \text{(equals for ideal cavity)} \quad \text{(Eq. S11)}$$

$$P_{th} \geq 23.2 \; \mu W$$

where $P_{th}$ is the pumping power at threshold, $\alpha = 0.075$ is the absorption coefficient of monolayer WSe$_2$, $\upsilon_{pump}$ is the pumping source frequency, $\upsilon_{lasing}$ and Q represent the frequency and quality factor of $TE_{1,17}$ mode lasing peak.

Finally, by combining some of our experimental parameters, we quantize the results that the pumping power should be at least larger than 23.2 μW to meet the quantum threshold condition. Therefore, the threshold pumping power in our PL experiment (Fig. S5b) is around 77 μW, which reveals that our device can truly satisfy the quantum threshold condition in room temperature.

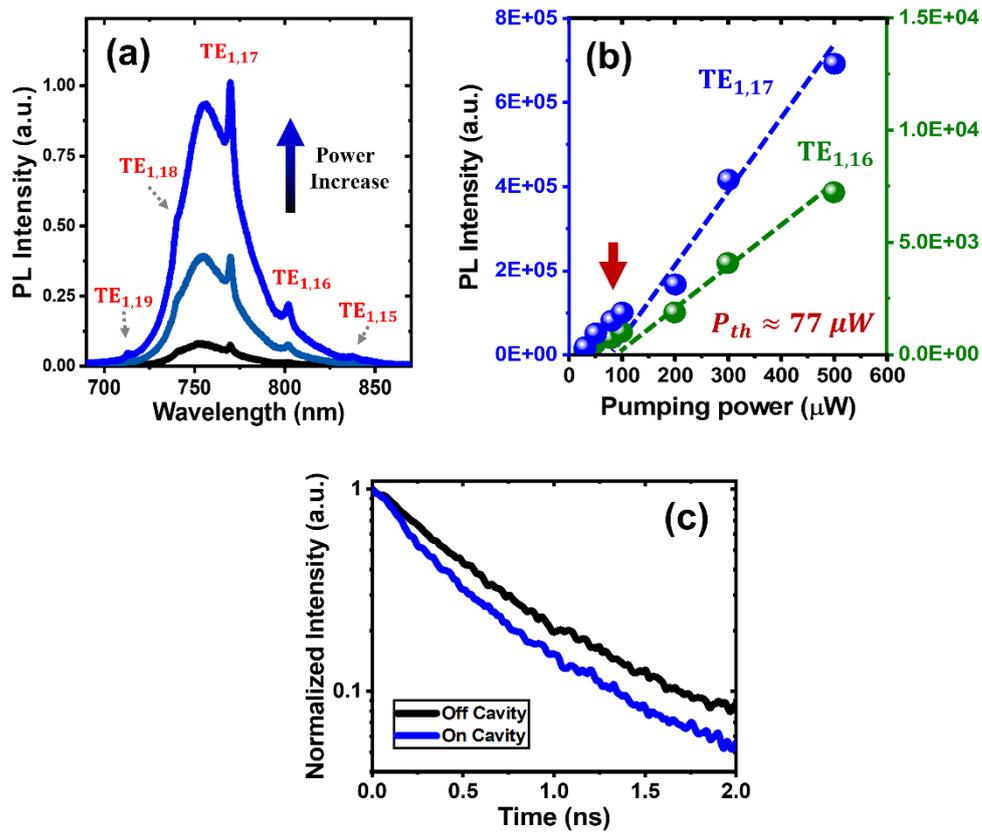

**Fig. S5 a** PL spectra of our $WSe_2$/microdisk laser device pumped by 532 nm laser with pumping power increasing. **b** Input–output curves of PL intensity extracted from $TE_{1,17}$ and $TE_{1,16}$ mode lasing peak as a function of pumping power. The pumping threshold is around 77 μW. **c** TRPL measurements of gain material $WSe_2$ on and off microdisk cavity under same pumping power and repetition rate.

## S6. Second-order coherence function $g^{(2)}(\tau)$ measurement setup

Fig. S6a illustrates the schematics of the optical setup of $g^{(2)}(\tau)$ measurement, which is known as a Hanbury Brown-Twiss (HBT) interferometer system. The EL/lasing signal generated from our $WSe_2$-microdisk cavity is collected by a 100× objective lens and split into two paths by a 50/50 beam splitter. The signals from different paths were then collected respectively by two photon-detection modules (PDM1&2) and analyzed by a Picoharp 300 time-correlated single photon counting (TCSPC) system. In addition to the EL measurements, PL measurements for $g^{(2)}(\tau)$ also revealed similar results below and above threshold in Fig. S6b and S6c, characterized the property of bunching and lasing, respectively.

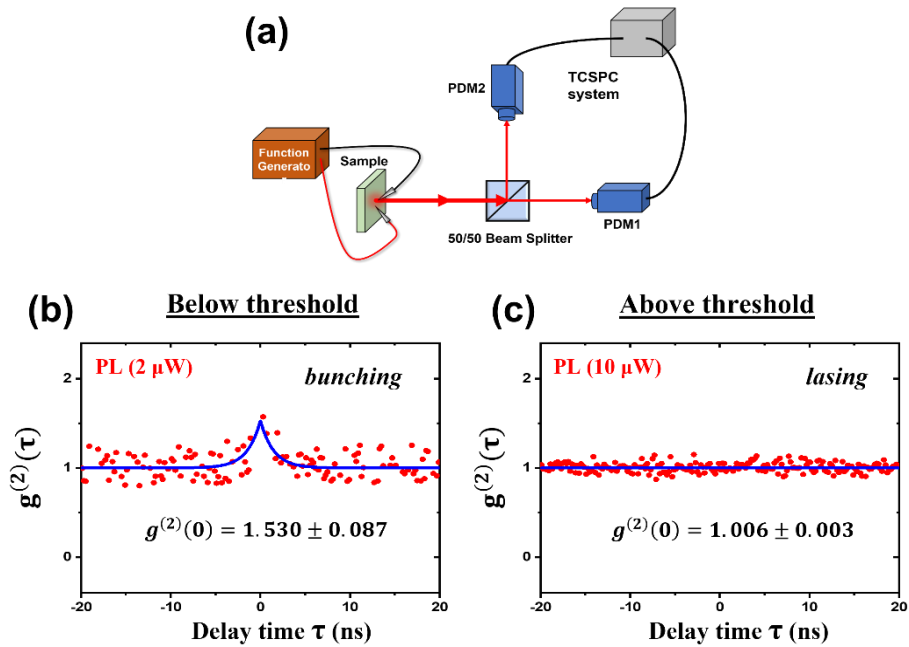

**Fig. S6 a** Schematic of the setup of EL $g^{(2)}(\tau)$ measurements. **b, c** Second-order coherence function PL measurements of the $WSe_2$ on microdisk device below and above lasing threshold. Red dots represent the raw data, and blue solid line is the corresponded fitted results from Siegert relation.

## S7. Prediction of electrode-gap distance (D) with simple effective RC model

To control carriers to recombine at the gap site for stronger coupling efficiency, it became really important to tune the electrode-gap distance D precisely. The gap width (G) is fixed at 0.4 µm due to the results in Fig. 4e, therefore the schematic diagram of our device is illustrated in Fig. S7a with several parameter marked. Referred to previous study, our device can be modeled into a simple effective RC circuit (Fig. S7b), where $R_{WSe2}$, $C_I$, $C_E$ and $C_M$ represents monolayered WSe$_2$ resistance, metal-WSe$_2$ interface capacitance, SiN$_x$ capacitance at electrode and SiN$_x$ capacitance at microdisk, respectively. To quantitively depict the circuit, classic parallel-pad capacitance formula was implied,

$$C(F) = \frac{\varepsilon_r \varepsilon_o A}{d} \qquad \text{(Eq. S12)}$$

where capacitance value is in unit of Farad (F), $\varepsilon_r$ is relative permittivity, $\varepsilon_0$ is vacuum permittivity, A is the pad's area and d is the vertical between two parallel pads. Therefore, three geometric capacitances in our devices' circuit can be calculated as follow,

$$C_I = \frac{1 \cdot (8.85 \times 10^{-12}) \cdot (5000 \ \mu m^2)}{(0.1 \ \text{nm})} = 442.5 \ (\text{pF})$$

$$C_E = \frac{7 \cdot (8.85 \times 10^{-12}) \cdot (5000 \ \mu m^2)}{(150 \ \text{nm})} = 2.065 \ (\text{pF})$$

$$C_M = \frac{7 \cdot (8.85 \times 10^{-12}) \cdot (\frac{10.76 \times 0.08}{10.76 + 0.08} \ \mu m^2)}{(150 \ \text{nm})} = 0.033 \ (\text{fF})$$

And the resistance of the WSe$_2$ ($R_{WSe2}$) material we used can be determined by a simple $I_{ds}$-$V_{ds}$ measurement across a 5 µm channel (Fig. S7c). The slope after Shockley barrier reveals approximately 12.1 GΩ of resistance, that is, 2.42 GΩ per µm. Thus, based on the length of WSe$_2$ we will across, $R_{WSe2}$ can be described as 2.42×(0.4+D) GΩ with vary D (Fig. S7d).

Herein, we considered only on the circuit that cross the electrode, WSe$_2$, air gap and microdisk (yellow part in Fig. S7c) to streamline into a simple effective AC supply RC circuit (Fig. S7d) with

$$C_{eff} = \frac{1}{\frac{1}{C_I}+\frac{1}{C_M}} = \frac{C_I \times C_M}{C_I + C_M} \approx C_M \ (C_I \gg C_M) \quad \text{(Eq. S13)}$$

Since, we are using a MHz-scale high frequency AC supply as source here, the charging time delay (RC delay) of capacitances become non-negligible. The voltage polarity change in time of capacitance can be written as,

$$V_c(t) = V_0(1 - e^{-\frac{t}{RC}}) \quad \text{(Eq. S14)}$$

where $V_c(t)$ is the bias voltage been charge in function of time, $V_0$ is the bias voltage with capacitance fully charged.

The voltage polarity will reach an effective inversion when t ≥ RC, that is, in our cases, when

$$\frac{1}{C_{eff}\ charging\ time\ t} = AC\ polar\ changing\ frequency \quad \text{(Eq. S15)}$$

➔ Alternative Positive/Negative charges will meet right near the capacitor interface, which is happened to be where the gap locates (Fig. S7d).

Therefore, to optimized the distance D for carriers to meet at the gap site,

$$\frac{1}{t} = \frac{1}{R_{WSe_2} \times C_{eff}} = 16\ MHz$$

$$\frac{1}{2.42\ (\frac{G\Omega}{\mu m}) \times (D + 0.4)(\mu m) \times 0.033(fF)} = 16\ (MHz)$$

$$D \approx 383\ (nm)$$

By simple RC model calculation, we can predict that when D equals to 383 nm, the alternative positive/negative charges will meet and recombine together right at the gap site for effective coupling with cavity mode.

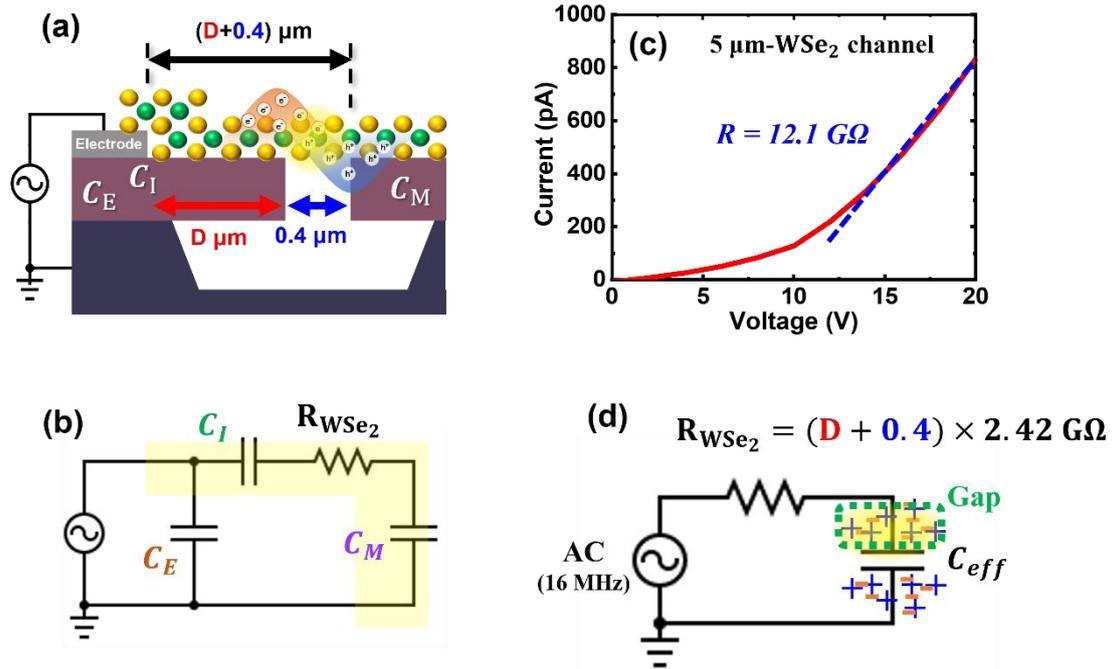

**Fig. S7 a** Schematic of the cross-section of our device with D and fixed 0.4 μm of G marked as the contact-gap distance and gap width, respectively. Three geometric capacitances $C_E$, $C_I$ and $C_M$ have also marked at their corresponded place. **b** Modelled effective RC circuit of our lasing device. **c** The measured $I_{ds}$-$V_{ds}$ curve of our monolayer WSe$_2$ material with 5μm's distance between source and drain contacts. **d** Simple AC-supply effective RC circuit depicting the D-dependent WSe$_2$ resistance and alternative positive/negative charges to recombine near the capacitance, right at gap site (Green dashed box).